\documentclass[usenatbib]{mn2e}
\usepackage{psfig}
\usepackage{times}

\def\solmas{{M$_\odot$~}}
\def\solmasp{{M$_\odot$}}
\def\tff{t$_{ff}$~}
\def\sims{$\sim$~}
\def\sfc{SFC~}
\def\sfcp{SFC}
\def\sfcs{SFCs~}
\def\sfcsp{SFCs}

\title[Unbound GMCs and OB associations] {Star formation in unbound giant
molecular clouds: \\ the origin of OB associations?} 

\author[P.C. Clark et al]
{Paul C. Clark$^1$ \thanks{E-mail: pcc@st-and.ac.uk} Ian A. Bonnell$^1$, Hans
Zinnecker$^2$ \&
Matthew R. Bate$^3$\\ $^1$ School of Physics \&
Astronomy, University of St Andrews, North Haugh, St Andrews, Fife, KY16 9SS. \\ 
$^2$ Astrophysikalisches Institut Potsdam, An der Sternwarte 16, D-14482
Potsdam, Germany \\
$^3$ School of Physics, University of Exeter, Stocker Road, Exeter, EX4 4QL}

\date{\today}

\begin{document}

\maketitle

\begin{abstract}

We investigate the formation of star clusters in an unbound GMC, where the
supporting kinetic energy is twice as large as the cloud's self-gravity. This
cloud manages to form a series of star clusters and disperse, all within
roughly 2 crossing times (10 Myr), supporting recent claims that star
formation is a rapid process. Simple assumptions about the nature of the star
formation occurring in the clusters allows us to place an estimate for the star
formation efficiency at about 5 to 10\%, consistent with observations. We also
propose that unbound clouds can act as a mechanism for forming OB associations.
The clusters that form in the cloud behave as OB subgroups. These
clusters are naturally expanding from one another due to unbound nature of the
flows that create them. The properties of the cloud we present here are are
consistent with those of classic OB associations.

\end{abstract}


\section{Introduction}
\label{intro}

Practically all star formation is thought to occur in clusters that are
embedded in giant molecular clouds (GMCs) \citep{Ladas2003}. This suggests that
a clustering environment, where multiple objects compete for a common gas
reservoir \citep{Zinnecker1982, Larson1992}, plays an important role in early
stages of protostellar evolution, such as dictating the form of the stellar
initial mass function (IMF) (\citealt{Bonnelletal1997};
\citealt{Bonnelletal2001b}). Furthermore, \citet{Elmegreen2000} has collected
observational evidence suggesting that star formation is a rapid process,
occurring on roughly the crossing time of the region at a variety of scales.
Not only does he propose that the star formation in a typical GMC occurs within
\sims 4 Myr (approximately the crossing time for standard GMC) but that the
cloud's dispersal occurs within a few crossing times, or $\la$ 10 Myr. 

The combined implication of these observations is that star formation occurs
quickly and in groups and that the sites of star formation disperse quickly. Our
proposal in this paper is that this is possible if GMCs are dynamically unbound
objects, with the internal turbulent energy greater than that of the cloud's
self-gravity. This follows from the work of \citet*{Semadenietal1995} who showed
that transient (unbound) GMC sized objects can be formed from flows in the ISM.
We also find that unbound GMCs may provide a natural mechanism for the creation
of OB associations, a notion first suggested by \citet{Ambart1958}.

In the rest of this first section we discuss the ideas behind rapid star
formation and the dynamical state of GMCs. We also include in this section a
discussion of OB associations. In section {\ref{setup}} we describe the details
of the simulation and section {\ref{evol}} follows the general evolution of the
GMC. In section {\ref{SFE}} we give estimates of the star formation efficiency
in the GMC based on some simple assumptions. In section {\ref{OB}} we highlight
the similarities between the simulation and the general structure in an OB
associations. A summary of the paper's main conclusions can be found in
section {\ref{finish}}.

\subsection{GMC Lifetimes and Rapid Star Formation}
\label{GMCS}

Until the last decade or so GMCs were generally believed to be long-lived
structures, with some estimates of ages reaching as high as $10^{8}$ Myr
\citep*{Solomonetal1979, Scovilleetal1979, ScovilleHersh1979}. It was generally
believed that the chemistry of turning atomic species into molecules would
require millions of years before an object like a GMC would be detectable via
its CO abundance \citep{Jura1975}. One also had the problem that the CO mass in
the galaxy, coupled with estimates of the star formation rate, suggested that
GMCs had to live for tens of millions of years if the star formation efficiency
was to remain at the observed level of a few percent \citep{ZuckermanEvans1974,
ZuckermanPalmer1974}. 

Recent observations of embedded clusters tend to suggest that the whole process
of star formation, including GMC formation and dispersal, occurs on roughly the
crossing time for the region \citep{Elmegreen2000}. Not only do most molecular
clouds in the solar neighbourhood contain signs of star formation in the form
of clusters, but the age determination of these clusters suggests they are very
young, typically less than 10Myr \citep{Hartmann2000}. This suggests that star
formation occurs quickly in GMCs after their formation. The fact that clusters
with ages greater than \sims 5 Myr are seldom associated with molecular gas,
suggests that clouds disperse quickly \citep*{Leisawizetal1989}.

In the original cloud lifetime proposition, it was assumed that all of the CO
observed in the galaxy was associated with molecular hydrogen involved in star
formation. We now realise that the vast majority of the gas that comprises a
GMC is never involved in the star formation process. In fact the star formation
efficiency in GMCs is only a few percent. The reason behind this lies with the
fact that little of the cloud is actually dense and bound enough to turn into
stars in the cloud's lifetime \citep{Padoan1995, Hartmann1998, Zinnecker2002}.
Also, if GMCs are short-lived features then there is little time for the more
tenuous parts of the cloud to get involved in the star formation via
accretion. 

The old GMC model also required the cloud to be supported and in virial
equilibrium, since that would permit them to remain as coherent structures for
as long as was necessary. This support pressure had to be in the form of
non-thermal kinetic energy, such as turbulence \citep{Larson1981}, since the
thermal energy component of these clouds is typically very small. To counteract
the gravity on the large scales however requires motions which are supersonic,
and it is known that these quickly damp in shocks (\citealt{Maclowetal1998};
\citealt*{Stoneetal1998}), even in the presence of magnetic fields. Thus the
bound GMC model requires some method of continually driving the turbulence on
the large scale. These driving mechanisms are not necessary in the short cloud
lifetime model, and there also is no need to assume that the clouds are in
virial equilibrium. GMCs can therefore exist in a variety of dynamical states.

\citet*{Heyeretal2001} have examined the stability of molecular clouds in the
outer galaxy and come to the conclusion that most clouds are indeed globally
unbound by their internal motions. They also point out the difficulty in
producing mass estimates (which a great number of papers on the subject of GMCs
pass over) and note that even mass estimates determined via CO measurements
(both $^{12}$CO and $^{13}$CO) assumes at some stage the cloud is bound.
Although they do find the clouds approach dynamical stability at large masses
($>10^{5}$\solmasp), there is still considerable scatter in the data,
suggesting that there is no typical dynamic state for GMCs. 

\citet*{Pringleetal2001} have shown that it might also be possible to build
GMCs by accumulating very low density hydrogen gas, already in a molecular
state. Their study came in response to the ideas presented by
\citet{Elmegreen2000}, in an attempt to provide a new mechanism for GMC
formation that can occur quickly. They point out that it is quite possible that
a large fraction of the interstellar medium may be in molecular form, but
either simply too low a density to be detectable by current methods or too far
away from illuminating sources. The GMCs are then formed from large scale
shocks, from spiral arm passage or feedback from high mass stars, such as winds
and supernovae. This cloud formation can occur within a few million years.
\citet{Pringleetal2001} also point out that GMCs are probably not in virial
equilibrium, and note that their wind-swept appearance suggests that they are
anything but.

The simulation that we present here draws on the above studies for motivation.
We assume that large scale flows are able to create an unbound GMC in a few
millions years. Instead of being contained by external forces (e.g.
\citealt{Heyeretal2001}), we assume that the cloud is free to expand into the
ISM. Thus the flows that created the cloud are assumed to have been used up in
its formation. Since the cloud is assumed to be short lived and not
quasi-static, there is no need for the internal turbulent energy, which will
dissipate on the crossing time, to be replenished \citep*{Paredesetal1999,
Elmegreen2000}.

\subsection{The Origin of OB Associations}
\label{originOB}

OB associations are historically identified simply as extended groups of OB
stars, having diameters of tens of parsecs \citep{Ambart1955}. Furthermore they
are rather more diffuse than open clusters, with the mass density of OB type
stars at $\sim 0.1$ \solmas pc$^{-3}$ \citep{Blaauw1964, Ambart1955,
Garmany1994, Ladas2003}. It was found that these associations contain
considerable substructure which are referred to as `OB subgoups'
\citep{Blaauw1964}. These subgroups are unbound from one another as was deduced
from their expansion about the centre of the region \citep{Blaauw1952}. Some
regions or `subgroups' are shown to be associated with molecular gas. In
general these regions are not coeval but can exhibit a spread of ages between
the subgroup population as large as 10 Myr \citep{Blaauw1964}. The fact that
OB associations are very young, with some of the subgroups possessing ages of
the order of a millions years, suggests that unbound nature of the subgroups
from one another is primordial.

The relationship between OB associations and other types of clusters found in
the galactic disc, such as open clusters and embedded clusters, is still rather
unclear. The OB associations do however have a classic theory regarding their
formation. \citet{ElmegreenLada1977} proposed that OB associations form via
triggering, prompted by the ionised regions produced by previous generations of
OB stars. In this manner, the star formation is self propagating, with one
generations of OB stars triggering the formation of the next. Since the shocked
layer in which the new group of OB stars forms is moving away from the older OB
stars, at a few kms$^{-1}$, the new group is unbound from its parent group. The
region then naturally has the dynamics of the observed OB groups. Motivation
came from observations of stars forming at the boundaries of molecular clouds
and HII, such as NGC7538, M17 and M8 (\citealt*{Habingetal1972};
\citealt{Ladaetal1976}).

The issue is complicated however when one considers the detailed stellar
population of OB associations \citep{Garmany1994, Brown2001}. In the self
propagating model, OB type stars form in the shocked layers where conditions
are naturally more suited to forming high mass stars. Low mass stars form
spontaneously in the rest of the cloud. Thus the model assumes a two step
formation process whereby low mass stars and high mass stars are formed by
different mechanisms and in physically separated locations. The IMF of the OB
associations however do not exhibit this feature and generally possess the
standard field star IMF, at least within the Salpeter range (e.g. Sco OB2,
\citealt{deGeus1992}; \citealt{PreibischZinnecker1999}). Since up to nearly 90\%
of star formation is thought to occur in embedded clusters, with a field star
IMF and primordial mass segregation (for a discussion see \citealt{Ladas2003}),
it may be that the formation of OB associations has more in common with
standard clustered star formation.

\citet*{BBV2003} and \citet*{BVB2004} have modelled cluster formation in a
turbulently supported cloud. They modelled a 1000\solmas molecular cloud that
was initially supported against collapse by a turbulent velocity field. It was
found that the dissipation of the large scale supersonic flows produced a
number of distinct subclusters. Each subcluster contains at the core a massive
star. The subclusters were mass segregated and each had a protostellar
population consistent with that of the observed field star IMF, both of which
are the result of competitive accretion. Since the cloud was initially bound,
even more so after the dissipation of the turbulent energy, the whole system of
subclusters are themselves bound to one another. They quickly merge
within roughly 0.5 Myr (roughly twice the free-fall time for the original
cloud). If this merging process was to occur on large scales, such as a whole
GMC, one would never be able to form OB associations. The massive stars at the
centres of the subclusters would find themselves in one large cluster.

Our proposal in this paper is that OB associations are just a series of
clusters that form in {\it{unbound}} GMCs. The expanding cloud produces a
series of clusters that are unbound from one another due to the fact that the
flows that form them are also unbound from one another. The clusters, which
become OB subgroups, then simply expand away from their mutual centre of mass
along with the gas from the cloud, rather than merge into a single cluster.
Thus only one star formation mechanism is at play here: clustered formation.
The OB association therefore will have the universally observed IMF.

\section{Details of the GMC Simulation}
\label{setup}

\begin{figure*}
\centerline{\psfig{figure=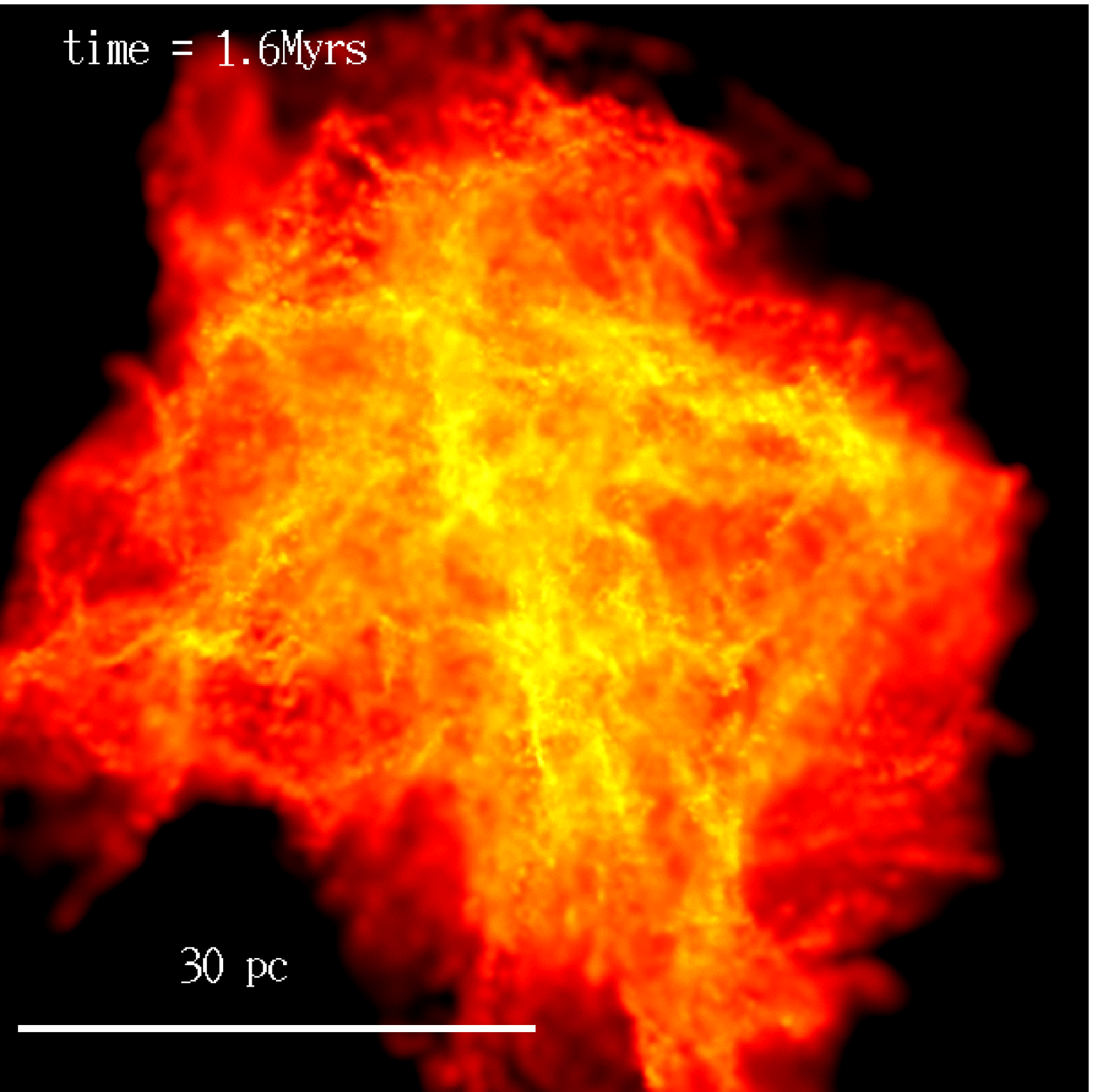,width=3.0truein,height=3.0truein}
\psfig{figure=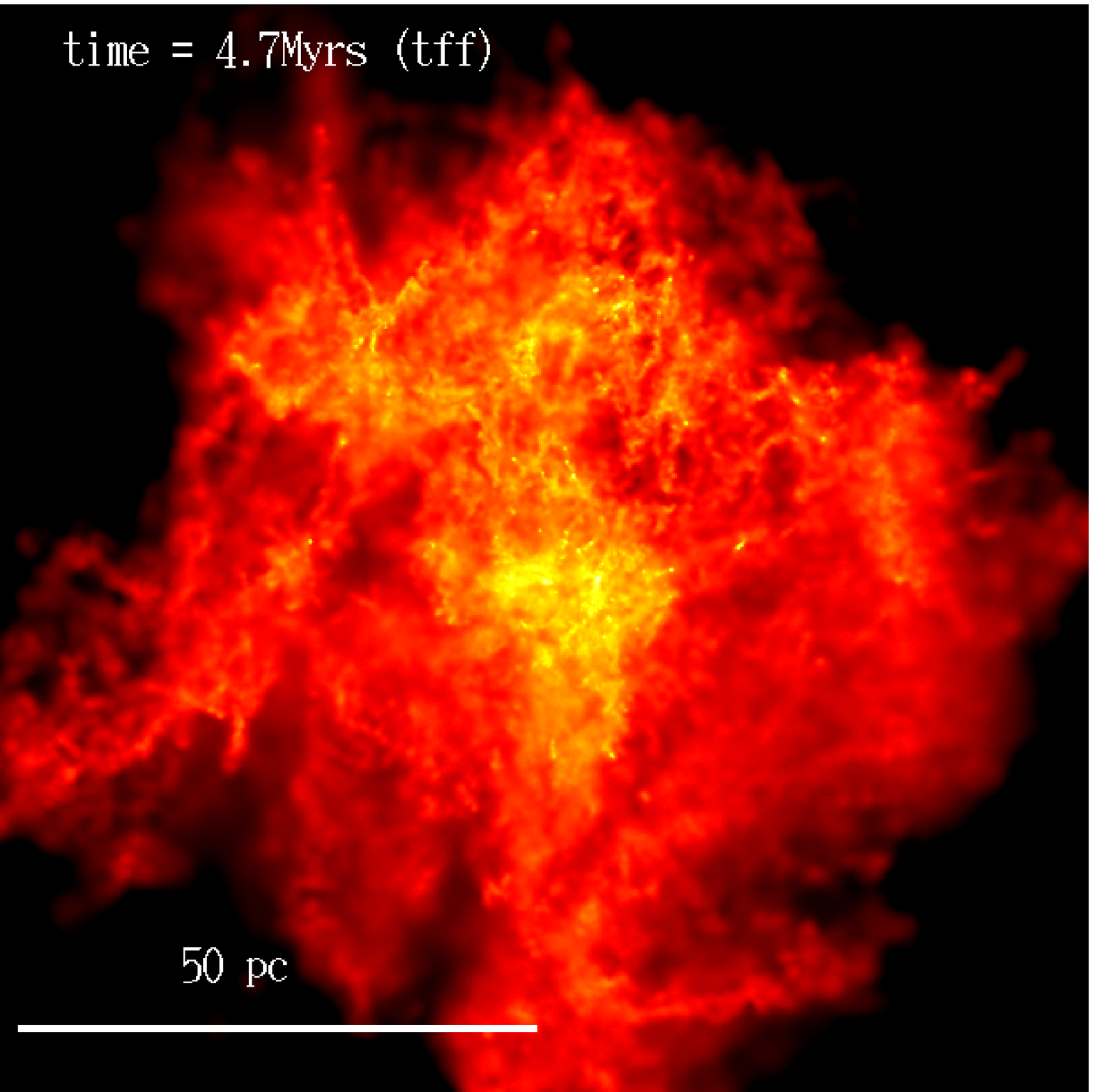,width=3.0truein,height=3.0truein}}
\centerline{\psfig{figure=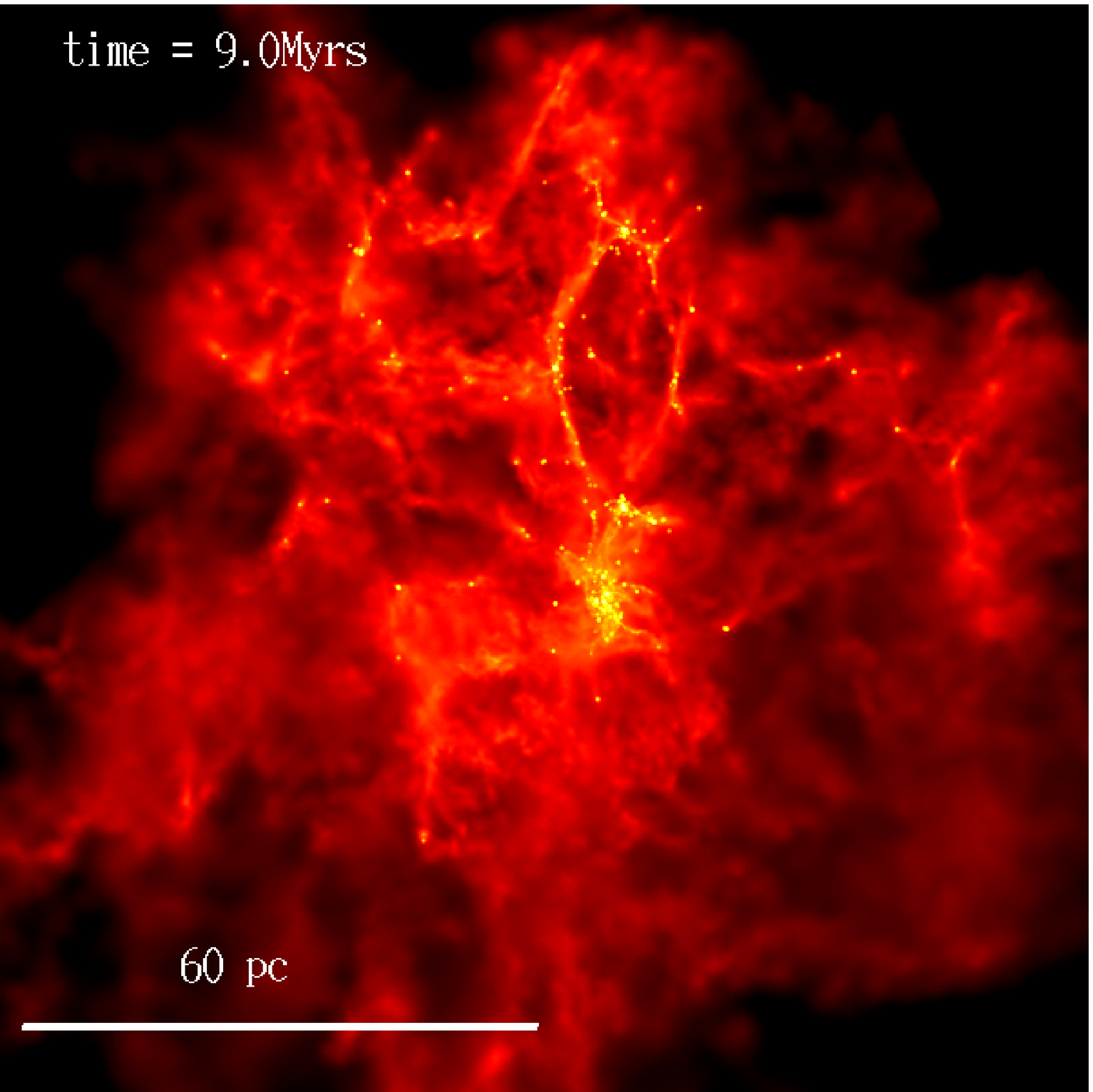,width=3.0truein,height=3.0truein}
\psfig{figure=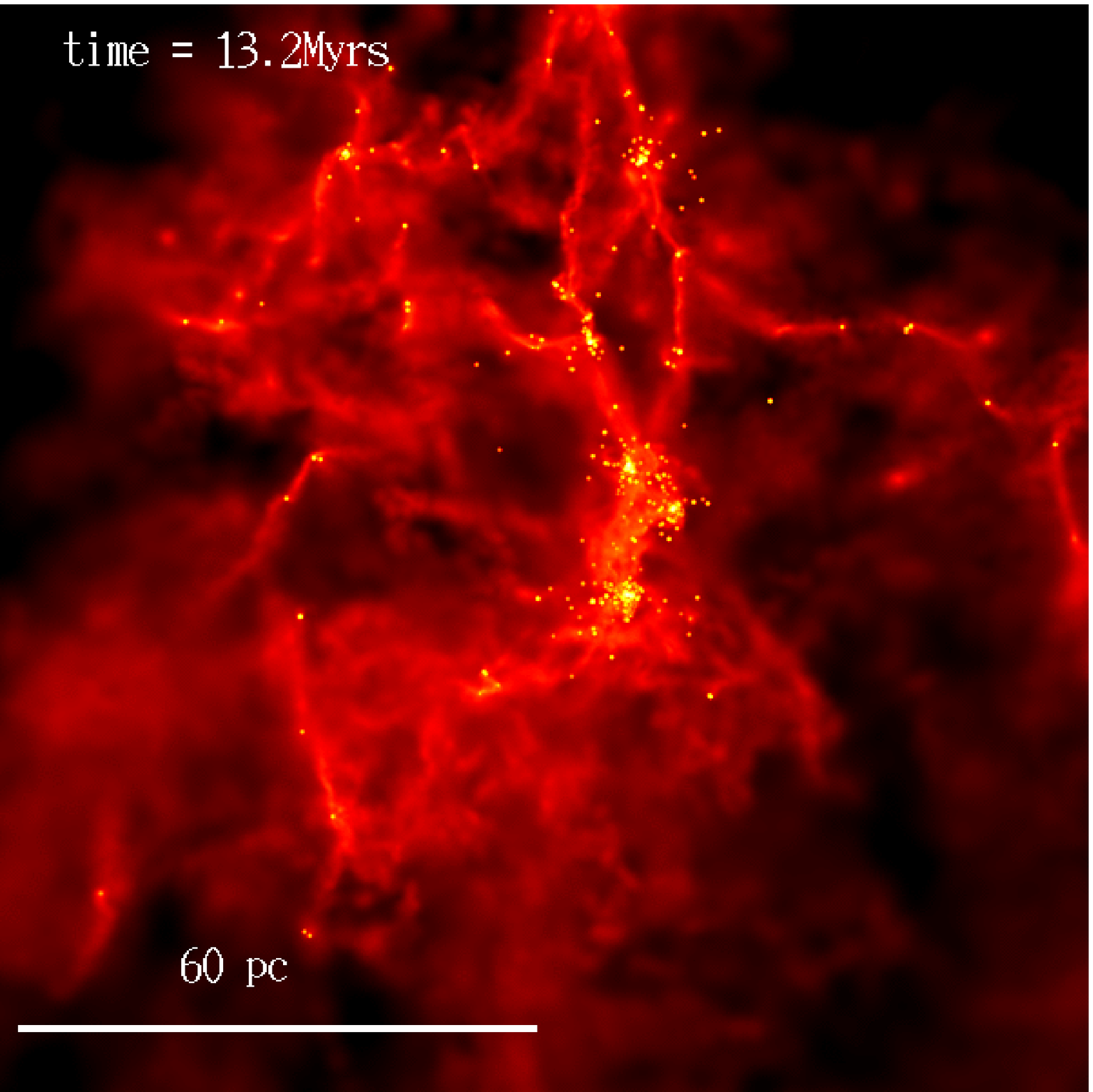,width=3.0truein,height=3.0truein}}
\caption{\label{piccies} The panels show column density images from the
simulations. The first panel shows the state of the gas after the crossing time
($t_{cr} \sim 1/3t_{ff}$), and the following panel shows the gas at the free
fall time. The remaining two images show how the cloud evolves to form a system of
clusters. The maximum column density plotted is 0.12, 1.21, 2.90 and 4.86 g
cm$^{-2}$ respectively and the minimum column density is $10^{-3}$ g cm$^{-2}$
in all the images.} \end{figure*}

 The fluid was modelled using the Lagrangian particle method of smoothed
particle hydrodynamics, or SPH \citep{Lucy1977, GingoldMonaghan1977}. The
smoothing lengths are variable in both time and space, with the constraint that
there must be roughly constant number of neighbours for each particle, which is
chosen to be roughly 50 (with a fluctuation from 30 to 70 neighbours). We use
the standard artificial viscosity suggested by \citet{GingoldMonaghan1983} with
$\alpha = 1$ and $\beta = 2$. Gravitational forces are calculated using dipole
and quadrupole moments obtained via a tree structure \citep{Benzetal1990}, which
is also used to construct particle neighbour lists. The code has been
parallelised by Bate using OpenMP and the simulation presented here was
performed on the UK Astrophysical Fluids Facility (UKAFF).

Our simulation starts with a uniform density sphere of molecular hydrogen of
radius 20pc with a mass of $1 \times 10^{5}$ \solmas. The gas is isothermal and
has a temperature of 10K. These numbers (mass, size and temperature) are
typical of those reported for GMCs in the solar neighbourhood
\citep{Blitz1991}. We model the gas with 500,000 SPH particles and are thus
able to accurately follow the formation of self-gravitating regions down to a
mass of 20\solmas \citep{BateBurkert1997, Whitworth1998}. The free fall time
associated with this cloud, the time taken for the unsupported gas to collapse
under gravity to a central point, is roughly 4.7 Myr. The cloud has an initial
Jeans mass of 30.4\solmasp. We do not include any feedback processes, such as
stellar winds and jets or the effects of massive stars such as ionisation
fronts and supernovae. 

To model the turbulence, we support the cloud with a Gaussian random velocity
field with a power spectrum of $P(k) \propto k^{-4}$ which is consistent with a
velocity field with a Larson-type relation of $\sigma \propto L^{0.5}$ where
$\sigma$ is the velocity dispersion and $L$ is the length scale of the region
\citep{MyersGammie1999}. At the beginning of the calculation the ratio of
gravitational to kinetic energy is 0.5 ($\mathrm{E_{kin} = 2 E_{grav}}$). We
stress that the turbulent kinetic energy is able to decay freely in this
simulation since we include no driving mechanism. The timescale for the energy
decay is the crossing time \citep{Maclowetal1998, Stoneetal1998} which for the
initial velocity field is $t_{cr} = $ 4.2 Myr, slightly less than the free
fall time.

The SPH code includes the modification by \citet{Bateetal1995}  which replaces
dense, self-gravitating, regions of the gas with point masses, or `sink
particles'. These sinks allow the code to model the dynamical evolution of
accreting protostars, without integration time steps becoming prohibitively
small. We set the sink particles to form at a density of 1000 times the initial
density, with a subsequent accretion radius of 0.17pc. When a particle finds
itself at the centre of a dense, bound and collapsing region it is turned into
a sink particle and its 50 to 100 neighbours are accreted onto it. With the
resolution used for this simulation the sink particles start with a mass of at
least 15\solmas before further accretion. Therefore we cannot think of these
point mass objects as `protostars', as was the case in \citet*{Bateetal2003},
but instead assume that they represent `proto-clusters'. To prevent the `sink
particles' behaving as point masses in gravitational interactions, we smooth
the sink-sink gravitational forces to a distance of $r_{min}$ = 0.2 pc in the
form $\mathrm{F_{ij} = -Gm_{i}m_{j}/(r_{ij} + r_{min})^{2}}$ between particles $i$ and
$j$.

In the analysis that follows, we discuss the properties of star formation
centres, or `\sfcsp'. These can either comprise of a single protocluster (or
sink particle) or a coherent group of protoclusters. To identify \sfcs where
more than one sink particle is involved, we make use of the mass segregation
that occurs naturally when the protoclusters interact in self gravitating
groups (see for example \citealt{BVB2004}). First we sort all the protoclusters
by mass. We then take the most massive protocluster and tag it and its fellow
protocluster neighbours within 0.5pc to be a members of `\sfc 1'. We then go
down the mass sorted list of protoclusters until we arrive at the next most
massive protocluster that has not been associated with \sfc 1. It becomes
tagged as being a member of `\sfc 2'. All the protoclusters within 0.5pc of
this protocluster are now tagged as being members of \sfc 2, unless they are
already members of \sfc 1. This process continues down the list of
protoclusters until all have been assigned membership to a \sfcp. There is thus
a resolution of 0.5pc which distinguishes one \sfc from another. We find that
the choice of 0.5 pc used in attributing memberships does not significantly
affect the \sfc population, since the protoclusters formed in the simulation
are either well separated (and thus in isolation) or exist in dense groups. The
radius of the \sfc is given by the radius of the furthest protocluster from the
centre of mass. If there is only one protocluster then the radius is simply the
accretion radius, which is 0.17pc.

One problem with trying to model turbulence in a numerical simulation of this
type is that it is not always possible to resolve the velocity structure at all
scales. Turbulence is assumed to be hierarchical, following a Larson-type
relation of $\sigma \propto L^{\alpha}$ \citep{Larson1981}. In SPH, while a
particle can have a kinetic energy based on its velocity, it can have no
internal velocity structure. As a result, the kinetic energy below a certain
mass scale ( actually the mass of an SPH particle and its neighbours) is not
included in the calculation. We therefore stress that our simulation is lacking
the kinetic energy that should be present at scales of less than \sims
20\solmasp. The details of how individual stars form are thus not available
from this simulation, and we must restrict ourselves to the large scale
properties of star formation and the formation of clusters.

\section{General Evolution} 
\label{evol}

\begin{figure*}
\centerline{\psfig{figure=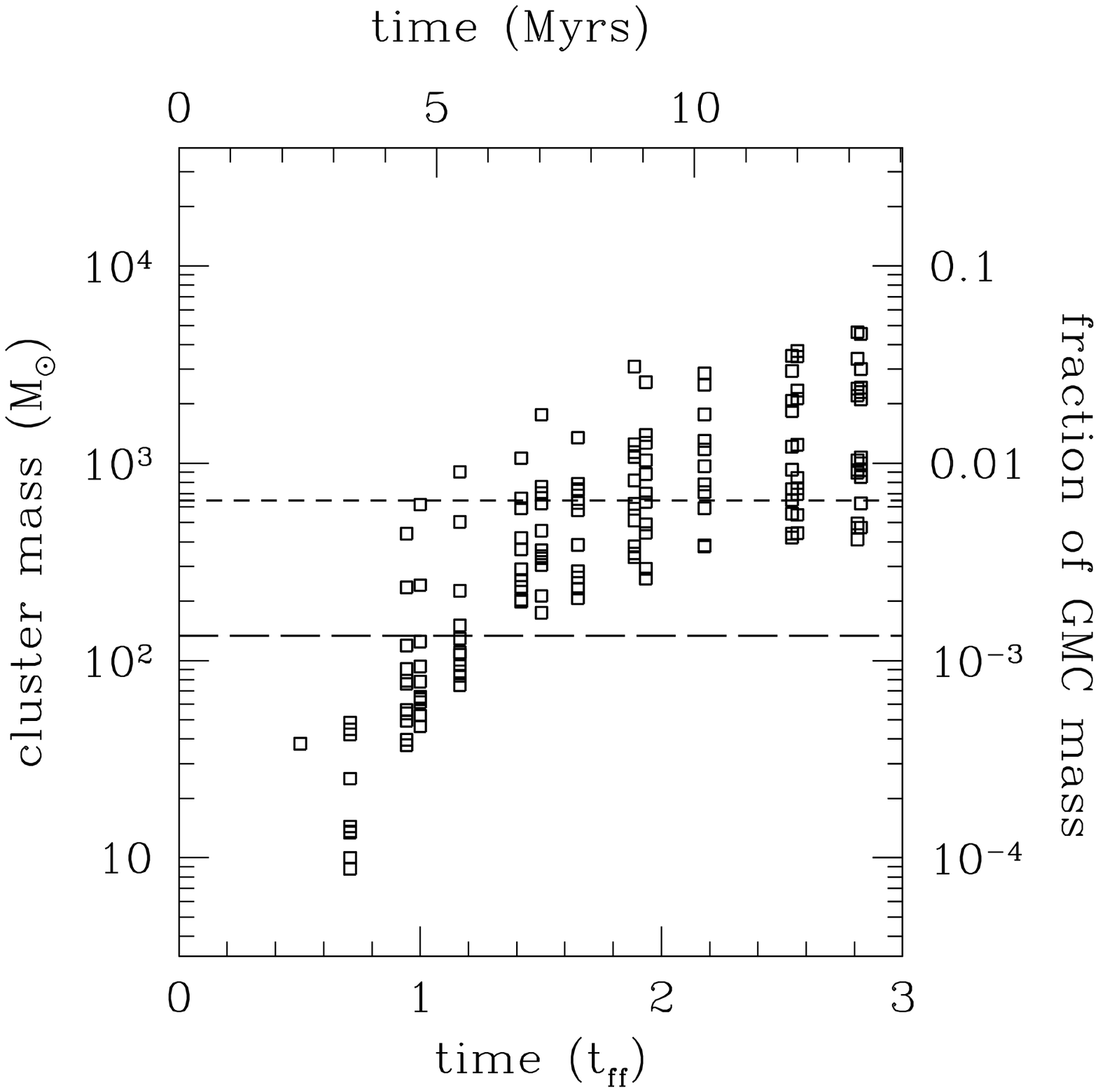,width=3.5truein,height=3.0truein}
\psfig{figure=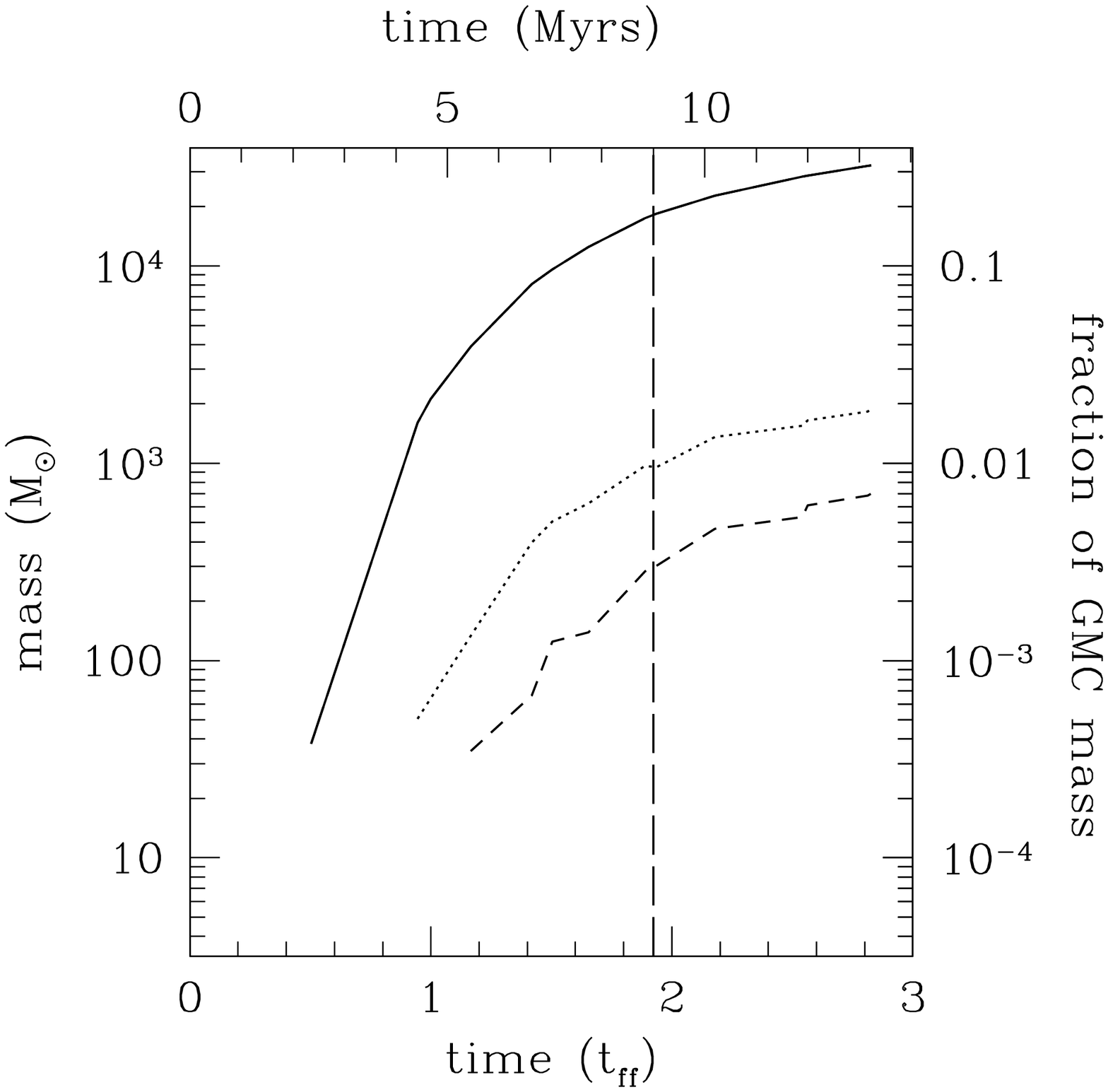,width=3.5truein,height=3.0truein}}
\caption{\label{bigclusters} Shown in the left hand panel are the masses of the
16 most massive \sfcs and how they evolve in time. The horizontal long-dashed
line gives the mass at which a \sfc can form a star of $>$ 10\solmas and the
short dashed line shows the mass needed by a \sfc to form a star of $>$
25\solmas. This assumes the IMF given in section {\ref{SFE}} along with a star
formation efficiency of 50\% in the \sfcs (see section {\ref{SFE}} for a
more detailed discussion). The solid line in the right hand panel shows the
mass contained in the \sfcs (gas particles + protocluster/sink particles).
The dotted line in the plot shows the mass in stars greater than 10\solmas
and the short-dashed line shows the mass in stars greater than 25\solmas. The
vertical long-dashed line denotes the point at which we estimate a supernova
event to occur (note that \tff = 4.7Myr).}
\end{figure*}

Figure {\ref{piccies}} shows column density images from different points during
the simulation and allows us to see clearly the evolution of the gas and
regions of star formation. We see from the figure that the structure of the
cloud changes remarkably quickly. It starts as a churning network of gaseous
filaments and within 10Myr (when the simulation was terminated) evolves into
an ensemble of distinct clusters, by which time the gas has lost much of its
early character. The fact that an unbound GMC can form stars and star clusters
reinforces the predictions made in \citet{ClarkBonnell2004}.

The point at which the first bound objects condense out of the unbound flows
occurs at roughly 2.4 Myr. This is roughly half the crossing time for the
region (although some authors use $t_{cr} = R/V$ instead of $t_{cr} = 2R/V$ as
is used here). This time is consistent with the kinetic energy dissipation rate
\citep{Maclowetal1998, Stoneetal1998} and the formation of a turbulently
dominated density structure \citep{Padoanetal2001}. 

Rather than simply discuss the individual protoclusters that form (the sink
particles) it makes sense here to discuss the bound groups of these
protoclusters as well, which we will simple refer to here as `star formation
centres' or \sfcsp. The formation of the \sfcs actually occurs very rapidly.
The mass of the 16 most massive of these centres is shown as a function of time
in figure {\ref{bigclusters}}. Within 5Myr (or \sims 2.5 Myr after the onset
of star formation) the 15 most massive \sfcs all have masses greater than about
100 \solmasp, and are beginning to get to a size where there is good
possibility of them forming massive stars (this will be discussed in section
{\ref{SFE}}).

A desirable feature of an initially unbound GMC is that cloud dispersal and
star formation are occurring simultaneously. This removes the necessity for
feedback mechanisms to disperse the cloud, or at the very least makes their
task much easier. The timescale for star formation is thus comparable to the
timescale for the cloud's dispersal. The dynamics of an unbound cloud is
therefore naturally in keeping with the recent observations that star formation
and cloud dispersal occur in a few crossing times. There is also the added
bonus that star formation efficiencies will be kept low, since most of the gas
around a protocluster clump will be unbound to it and moving away. This
prevents the material getting involved in the accretion once a \sfc starts
to form.

\begin{figure}  \psfig{figure=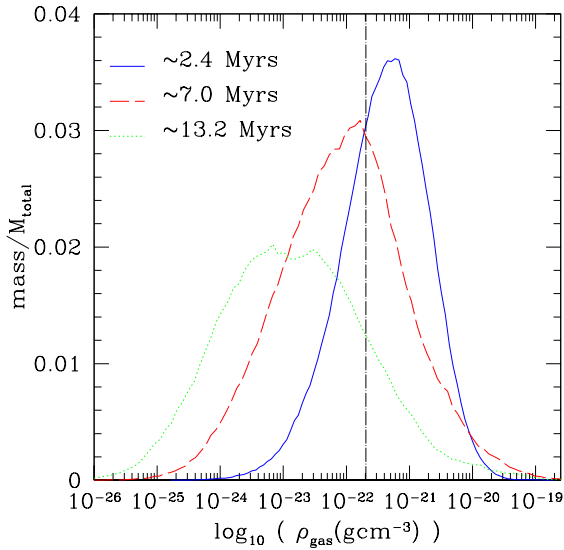,width=3.0truein,height=3.0truein}
\caption{\label{rhodist} The plot shows the density distribution of the gas at
three points in the simulation. The solid line follows the gas density just
before the formation of the first protocluster, at 2.4 Myr (or \sims
$\frac{1}{2}$ t$_{cr}$). The dashed and dotted lines show the density
distribution of the gas at a time of \sims 7 Myr and \sims 13.2 Myr
respectively. The vertical dot dashed line shows the density of the gas at the
beginning of the simulation. Note that the simulation only evolves a gas of
pure molecular hydrogen.}  \end{figure}

Figure {\ref{rhodist}} shows the density distribution of the gas at three
points in the simulation. The vertical dot-dashed line marks the original
density of the cloud. Just before the first protocluster forms at 2.4 Myr we
see that the most common density is roughly $7 \times 10^{-22}$ gcm$^{-3}$ (the
solid line curve), an order of magnitude higher than at the start of the
simulation. Note however that very little material at this point is as dense as
$7 \times 10^{-21}$ gcm$^{-3}$, showing that the turbulence does not allow much
material to get up to typical star forming densities \citep*{Falgaroneetal1991,
Padoan1995, Zinnecker2002}. 

After 7 Myr the peak in the distribution falls back to roughly the starting
density, however there is much more spread in the distribution. This spread is
controlled by two mechanisms. The high density tail increases as the \sfcs grow
by accretion and the subsequent rise in the potential energy. This causes yet
more material to fall into the star forming regions. The low density tail
increases since the cloud is freely expanding. By 13 Myr, only $\sim 3 t_{cr}$,
we see that the majority of the gas has fallen to very low densities. By this
point it is unlikely that observations of such a cloud would reveal much in the
way of molecular gas and would instead only be visible as HI. The cloud can now
be assumed to be `dispersed'. Even if the GMC fails to be a site of massive
star formation, the dispersal would still occur on a timescale consistent with
Elmegreen's (2000) observations.

Note also from figure {\ref{piccies}} that the cloud contains cavities and
dense regions of star formation. These are created in the simulation purely by
the turbulence. This type of structure in star forming clouds is often
attributed to the effects of high mass stellar feedback, such as winds and
supernovae, and is thought to be the trigger for star formation in the region
(e.g. \citealt{ElmegreenLada1977}) Instead, we realise that turbulence can
mimick these effects. Furthermore the cavities in the simulation would be
easily ionised by any high mass stars that form in the \sfcs
\citep{Daleetal2004}. We would then have a series of \sfcs separated by a
region of HII gas, just as is found in the classic picture of triggered star
formation.

\section{The Formation of Stars and Expected Efficiency}
\label{SFE}

In this section we use some simple assumptions about the star formation that
occurs in the \sfcs to determine the numbers of high mass stars and the star
formation efficiency that one might expect from the simulation. It is still
beyond the capabilities of current computational resources to model the details
of how individual stars form in a body of gas as large as a GMC. In the
simulation presented here we cannot model any gas dynamics below the 20\solmas
scale. We can however give the reader a feel for the star formation that is
present by using the results of previous simulations, along with some
assumptions about the star formation efficiency and the form of the IMF.

It has been shown from numerical simulations that star formation occurs on
roughly the local crossing time for the turbulence when the region is
dynamically bound \citep{Bateetal2003, Klessen2001}. On the small scales such as
those represented by our protoclusters, \sims 0.1pc, the crossing time is of the
order $10^{5}$ years. We can therefore assume that all of our protoclusters form
stars and that the star formation in our protoclusters takes place quickly,
rapidly enough to be regarded here as instantaneous compared to the evolution
of the whole GMC.

The simulation presented also has no method of incorporating feedback into the
GMC model. As is shown in figure {\ref{bigclusters}} in the right hand plot,
the mass accreted into the \sfcs gradually increases as the simulation
progresses. At the point where the simulation is terminated, 30\% of the GMC
has been accreted by the protoclusters. It is unlikely that this value is
representative of how much mass would actually be involved in the star
formation by this time, since feedback mechanisms such as ionisation, winds and
supernovae would seriously alter the amount of gas that would be available for
accretion into the \sfcsp. What is needed is an estimate of when one would
expect the star formation process to be halted by feedback mechanisms. This
requires some knowledge of the star formation taking place within the \sfcsp.

We have already pointed out that the protoclusters in the simulation group into
large \sfcsp. From now on in the paper we will use the details of these
regions, rather than the individual protoclusters, to assess the nature of the
star formation in the GMC. Table {\ref{clusterinfo}} gives the details of the
\sfcs after 9 Myr. The masses quoted for the \sfcs in table {\ref{clusterinfo}}
includes all particles (SPH and protoclusters) that fall within the radius of
the region. The gas particle component is however quite small, generally less
than 20\%.

Although the star formation efficiency of GMCs is thought to be in the range of
1 to 10\%, at the cluster level it is thought to be about 20 to 50\% depending
on the region (for a discussion we point the reader to \citealt{Ladas2003} and
\citealt{Kroupa2001}). In this paper we assume that the star formation
efficiency in our \sfcs is 50\%, but will include a discussion about the case
in which 100\% of the mass is turned into stars. The assumed efficiency here is
high but this is deliberate since it actually assumes as little as possible
about the effect that the feedback mechanisms from the young stars are having
on the accretion processes in the \sfcp. We will also assume that the IMF of
the stellar population in the \sfcs follow a two step power law form, $dN
\propto m^{-\alpha} dm$, with $\alpha$ = 1.5 for $0.08 < m/M_{\odot} \le 0.5$
and $\alpha$ = 2.35 \citep{Salpeter1955} for $0.5 < m/M_{\odot} \le 100 $. This
IMF, in conjunction with our assumption that 50\% of the mass of the \sfcs is
turned into stars, allows us to estimate the stellar population produced by the
simulation.

As already mentioned in the previous section, figure {\ref{bigclusters}} shows
in the left hand plot how the mass of the 15 largest \sfcs evolves with time.
The horizontal lines mark the point at which high mass stars can form. From our
IMF model, 15\% of the mass should be contained in stars with masses greater
that 10\solmasp. Thus a 10\solmas star will be present provided that there is
10/0.15 =67\solmas in the stellar population. Applying our assumed star
formation efficiency of 50\%, the \sfcs must therefore have a mass of
134\solmas if they are to harbour a 10\solmas star. The horizontal long-dashed
line in the figure denotes the point at which the \sfcs achieve this mass.
Doing the same for 25\solmas stars, which should comprise 7.7\% of the stellar
mass in our chosen IMF, we find that the \sfcs need to contain $25/(0.077
\times 0.5)$ = 650\solmas if they are to contain a 25\solmas star. This is
represented by the horizontal short-dashed line in the figure.

\begin{figure*} \centerline{
\psfig{figure=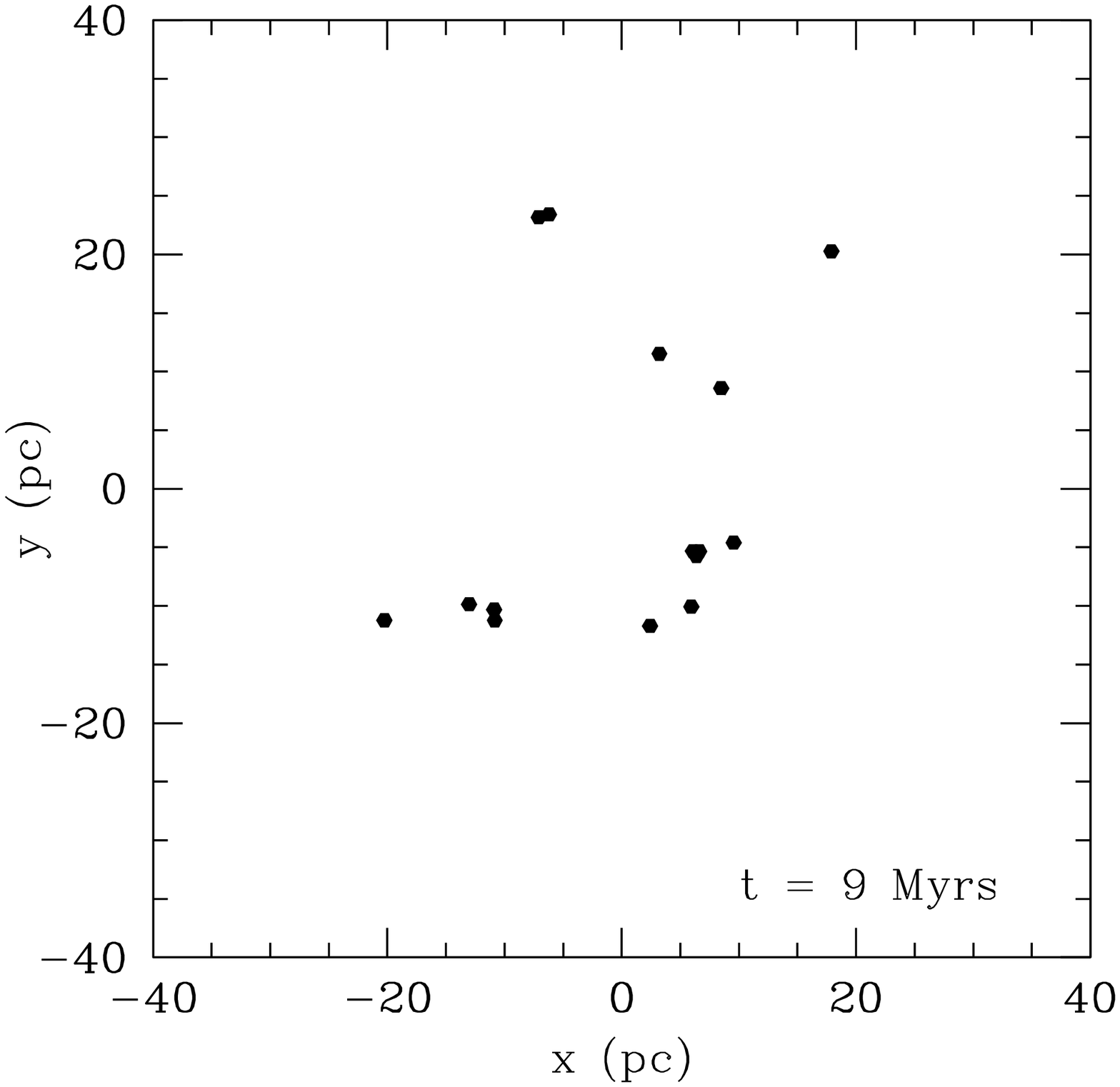,width=3.0truein,height=3.0truein}
\psfig{figure=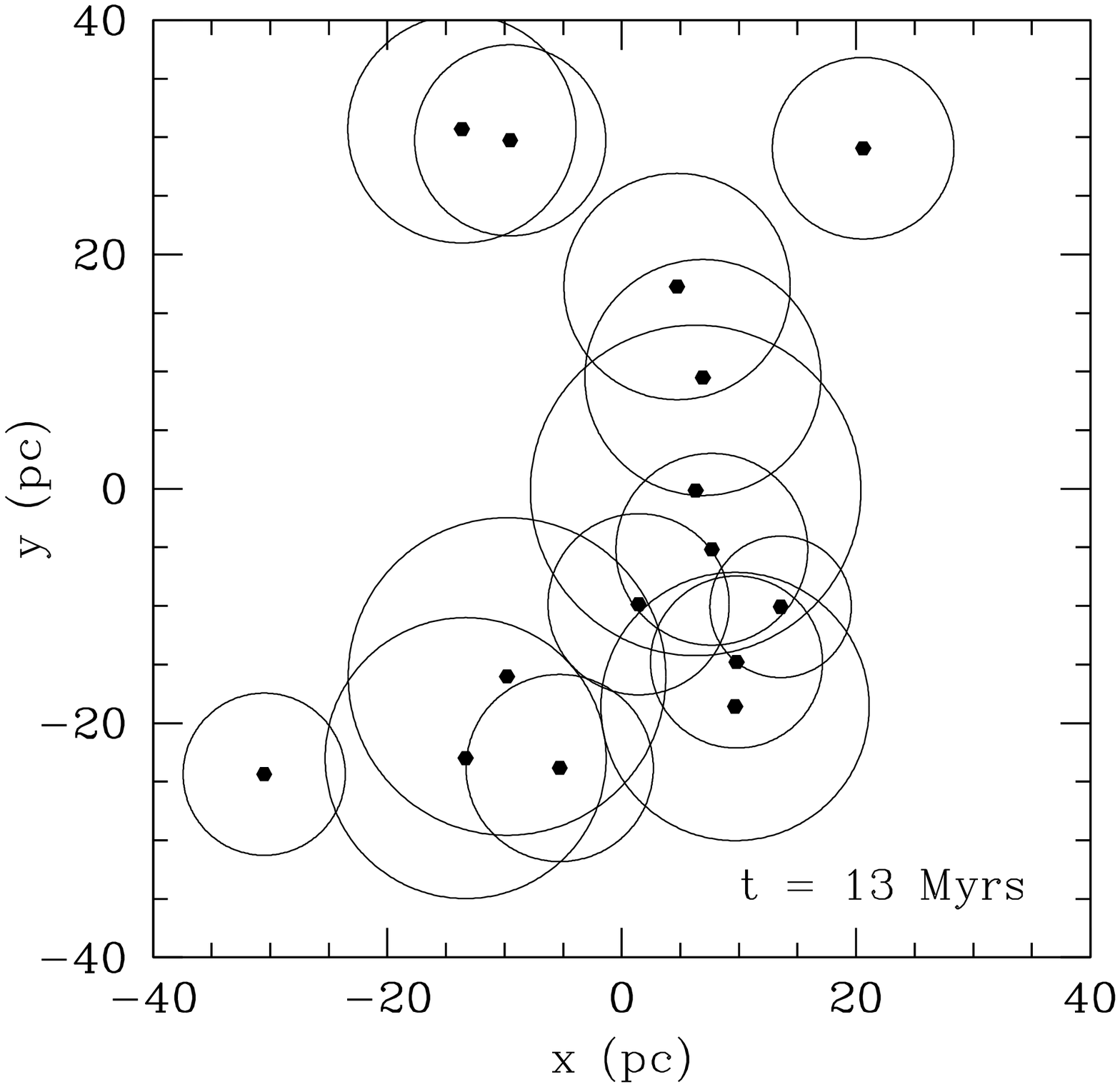,width=3.0truein,height=3.0truein}}
\caption{\label{OBcluster} Plotted are the positions of the \sfcs large enough,
by t = 9 Myr, to contain stars of mass greater than 10\solmasp. The plot on the
left shows the positions of the \sfcs at t = 9Myr and the plot on the right
shows how the \sfcs will be positioned after 13Myr, assuming that a SN event
expels the GMC gas and the \sfcs continue on their original path. The circles
show the size of the \sfcs after the 4 Myr period, assuming they expand with
their internal velocity dispersion. See section {\ref{OB}} for a discussion.} 
\end{figure*}

From our simple assumptions about the small scale efficiency and the form of
the IMF, we can estimate at what point in the simulation the star formation
process will be disrupted by feedback mechanisms. From figure
{\ref{bigclusters}}, we can estimate that the formation of 10\solmas stars
would occur at about 0.8\tff (or at \sims 4 Myr). A star of mass 25\solmas
would form after \sims 1.1\tff (or \sims 5 Myr). Since the mass of the \sfcs is
increasing fairly rapidly at this point, stars with even higher masses would be
expected to be present shortly after this, within 0.5 Myr or so. It would thus
appear that the GMC is able to get enough mass into the \sfcs for them to be
able to form a full stellar population within about 1 Myr. This is consistent
with the observations of the small age spread in the stellar population of the
Orion cluster \citep{Hillenbrandetal2001}.

Very rapidly after the first stars form we see that 10\solmas objects will be
present. This means that shortly after their formation, \sfcs are going to
contain ionising sources. Such stars are commonly suggested to be responsible
for controlling the star formation efficiency by expelling the gas from the
cluster in which they form (such as our \sfcsp), thus preventing the
protostellar population from accreting or preventing new stars from forming.
However \citet{Daleetal2004} have noted that the ionisation from these stars
does not appear to significantly affect the accretion rate in the clusters. The
clumpy/fractal nature of the gas at the centre of the cluster where the OB
type stars are situated acts to shield vast regions of the cluster from
ionisation. Rather than pushing through the dense material, the ionising
photons just find the path with the least resistance out of the cluster. This
is low density gas which would not normally be associated with protostellar
accretion in the first place. Similarly, the gas structure may also prevent the
winds from OB stars expelling gas from the cluster. It has been suggested that
winds are able to escape via the fractal holes, without imparting much momentum
to the dense regions \citep{Henning1989}.

It is therefore not clear if ionisation or winds will be able to expel the gas
from cluster, thus halting the star formation process. One mechanism that
certainly will produce the desired effect is a supernova explosion. In fact it
has been estimated that these events will not only remove the gas from a
cluster, but also be able to disperse the natal GMC. Thus a high mass star's
death will definitely mark the end of the star formation period in our cloud.
Stars with masses greater than 25\solmas have very short main-sequence
lifetimes, of about 3-5 Myr, and we see from the figure that they form at
about \sims 5 Myr after GMC formation. If we assume that a supernova event
will occur at about \sims 4 Myr after the formation of the very high mass
stars, then we estimate the first supernova event to occur at about 9 Myr or
when the OB stars are \sims 4 Myr old.

Assuming the supernova event will halt the star formation, we can now get an
estimate of the star formation efficiency in the GMC. The vertical dashed line
in figure {\ref{bigclusters}} denotes the point at which we might see the first
SN event. At this time, \sims 0.1 to 0.2 of the GMC's mass is contained in the
\sfcsp. However we have assumed up until now that the efficiency in the \sfcs
is not 100\% but 50\%, therefore our estimate of the star formation efficiency
in the GMC is roughly 5 to 10\%. This is easily comparable to the expected
efficiencies in GMCs by Elmegreen's (2000) rapid cloud formation/dispersal
model.

The above analysis relied on a lot of assumptions about the nature of the star
formation in the \sfcsp. In particular, it is guilty of invoking a star
formation efficiency in the \sfcs in order to determine the star formation
efficiency of the cloud: one could argue that this is not entirely
self-consistent. Here we redo the above analysis but without the \sfc
efficiency assumption. If all the mass in the \sfcs is used in forming a
stellar population, then the mass required by a \sfc before a 25\solmas star
forms is 25/0.077 = 325\solmas. The first \sfc to achieve this mass does so at
about 4 Myr, only 1 Myr less than our previous estimate. Thus we can
predict the supernova to occur at \sims 8 Myr. At this point in the simulation
there is about 7 - 8\% of the cloud incorporated in the \sfcsp. Thus the
expected star formation efficiency is still less than 10\%, suggesting that our
analysis is not heavily dependent on the assumptions. 

\begin{figure}  \psfig{figure=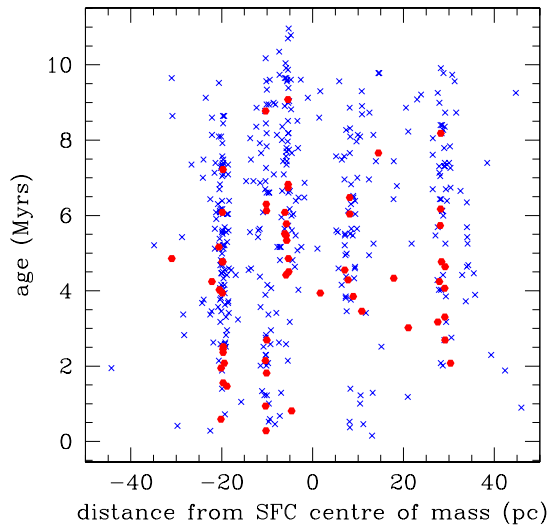,width=3.0truein,height=3.0truein}
\caption{\label{ages} The figure shows the age of the protoclusters (note, not
the \sfcsp but the individual `sink particles') as a function of distance. The
ages are shown at a time of 13 Myr after the cloud formation. The distance
shown is the y component from the centre of mass of all the protoclusters. The
crosses denote the age based on when the protoclusters form and the hexagons
points denotes the age based on when an individual protocluster can harbour a
star of greater than 10 \solmasp. See section \ref{OB} for more details. It is
clear from the distribution of points here that there is no discernable age
spread with position in the region. The star forming centers are essentially
coeval.} \end{figure}

\begin{table} 
\caption{\label{clusterinfo} The table gives the properties of all the star
formation centres (\sfcsp) that would be expected to contain massive stars by t
= 9 Myr (see section {\ref{SFE}} for a discussion of this). $\Delta$V is the
internal \sfc velocity dispersion and assumes the region is in virial
equilibrium such that $\mathrm{\Delta V  = (GM_{\sfcp}/R)^{1/2}}$. The crossing
time is then calculated from $\mathrm{t_{cr} = 2R/\Delta V}$. Note that
$\mathrm{M_{\sfcp}}$ is the total (gas + stars) mass enclosed within R.}
\begin{tabular}{c|c|c|c|c}
\sfc No. & $\mathrm{M_{\sfcp}}$ & $\Delta$V & R & $\mathrm{t_{cr}}$ \\ 
& (\solmas) & (km s$^{-1}$) & pc & Myr \\
\hline \hline
1  & 1763 & 3.27 &  0.71 & 0.42 \\
2  & 761  & 2.68 &  0.45 & 0.33 \\
3  & 706  & 3.24 &  0.29 & 0.17 \\
4  & 624  & 2.24 &  0.53 & 0.46 \\
5  & 455  & 2.37 &  0.35 & 0.29 \\
6  & 362  & 1.77 &  0.50 & 0.55 \\
7  & 338  & 1.77 &  0.46 & 0.51 \\
8  & 329  & 2.88 &  0.17 & 0.12 \\
9  & 305  & 1.68 &  0.46 & 0.54 \\
10 & 212  & 2.31 &  0.17 & 0.14 \\
11 & 174  & 1.37 &  0.39 & 0.56 \\
12 & 151  & 1.95 &  0.17 & 0.17 \\
13 & 150  & 1.95 &  0.17 & 0.17 \\
14 & 144  & 1.91 &  0.17 & 0.17 \\
15 & 141  & 1.63 &  0.23 & 0.27 \\
16 & 140  & 1.39 &  0.31 & 0.44 \\
\hline \end{tabular}  
\end{table}

\section{Dynamical Evolution of the Clusters and their relation to OB
associations} 
\label{OB}

As already discussed, the simulation produces a series of star formation
centres (\sfcsp). Of these regions, 16 of them (see table {\ref{clusterinfo}})
are massive enough to contain a star of greater than 10 \solmas by the time at
which we estimate a SN explosion will destroy the GMC. From figure
{\ref{piccies}} we see that these \sfcs are expanding as a group away from one
another (in fact this is true of the GMC structure in general). Furthermore,
the distance between \sfcs is roughly 10pc after about 13Myr. Thus the group
of clusters have the appearance of an OB association, with the individual \sfcs
being OB subgroups that are expanding about some common point. In this section
we examine the properties of the \sfcs and compare them to the observations of
OB associations.

In figure {\ref{OBcluster}}, the left hand panel shows the positions of the
\sfcs that are large enough to contain stars greater than 10\solmasp at t =
9Myr, assuming a star formation efficiency of 50\% and the IMF presented in
section {\ref{SFE}}. Their positions are plotted at the time we estimate the SN
event to start expelling gas from the cloud. We now assume that the SN event
removes the gas from the GMC quickly enough such that the motions of objects
have no time to adjust to the change in potential. We assume that this is true
both at the scale of the \sfc motions and at the smaller scale of the stellar
motions inside the \sfcsp. The right hand panel shows the positions of the
\sfcs after a further 4 Myr, i.e. at t = 13 Myr, assuming that they continue
on the path they had before the gas expulsion. The circles denote the size of
the \sfcs at t = 13 Myr, which have been evaluated from $\rm{r = r_{SN} +
\Delta V \times 4}$ Myr, where $\rm{\Delta V}$ is the region's internal
velocity dispersion and $\rm{r_{SN}}$ is the radius of the \sfcs at the point
when the SN event occurs (i.e. at \sims 9 Myr). This assumes that the star
forming regions will disperse at roughly their internal velocity dispersion
once the gas has been expelled.

Figure {\ref{OBcluster}} clearly shows that the \sfcs are expanding away from
one another. By determining the average radius of the \sfcs from their common
centre of mass, both at t = 9 Myr and at t = 13 Myr, we can get an estimate
of their expansion velocity from their dynamical centre. At t = 9 Myr the
average distance of the \sfcs from their centre of mass is 18.4pc, while after
13 Myr it is 25pc. This corresponds to an expansion velocity (that is a
3-dimensional velocity) for the \sfcs of 1.5kms$^{-1}$. We compare this, for
example, to the observed expansion of the OB subgroups in Per OB which is
roughly 2kms$^{-1}$ \citep{Fredrick1956, Blaauw1964}. We also see that if the
\sfcs themselves are able to expand after the gas expulsion, they become an
extended distribution of stars after only 13 Mys, as is shown by the circles in
the figure.

The stellar mass density is another important feature of OB associations.
Generally, OB associations have a density of OB stars of roughly 0.1\solmas
pc$^{-3}$ (see the introduction for references). In this simulation at t = 9
Myr, the density of OB type stars is 0.16\solmas pc$^{-3}$, and the mass
density of stars $\ge$ 25\solmas is 0.1\solmas pc$^{-3}$. This is calculated by
taking the total mass in the \sfcs of stars of greater than the required mass
type and dividing by the volume of region containing all the \sfcs with
these types of stars. After the system has had time to evolve for 4 Myr, the
densities are 0.06 and 0.04\solmas pc$^{-3}$ for the OB type stars and those
with masses $\ge$ 25\solmas respectively. Note these figures are based on the
\sfcs having a star formation efficiency of 50\% and containing the IMF of
stellar objects that was presented in section {\ref{SFE}}.

We can compare this to the density of high mass stars in the \sfcs at the point
of the SN explosion. If we take the largest \sfc, with mass 1763\solmas and
radius of 0.7pc, and assume again that 50\% of this is contained in stars. Then
the total mass in stars of mass greater than 10\solmas is $1763 \times 0.15 / 2
=$ 132\solmasp. The density of massive stars is then $132/0.7^{3} = 384$
\solmasp pc$^{-3}$. 

The turbulent flows are thus able to create a series of star forming regions
that have roughly the same properties as those found in OB associations. Since
the regions (the \sfcsp) are formed within large flows, the stars that form will
have roughly the same motion as the gas stream that formed them, potentially
explaining why \citet{Blaauw1991} finds that the gas surrounding OB subgroups
is generally moving with the group. 

Observations of some OB associations also indicate distinct age spreads between
their subgroups. This generally takes the form of an age progression from one
side of the association to the next. The ages are generally derived from where
the very high mass stars turn off the main sequence, which is a much more
reliable method than using pre-main-sequence (PMS) tracks of low mass objects.
Also the high mass end is normally the only part of the mass spectrum that is
well established in OB associations. This age spread has been the motivation
behind the triggered sequential star formation model developed by
\citet{ElmegreenLada1977}, which is in turn motivated by the observations of
\citet{Blaauw1964}. However, we note here that the Orion OB association
exhibits no discernible age progression in the subgroups \citep{Brownetal1999}.

Does our simulation show a convincing age spread between the subgroups/\sfcsp?
In figure {\ref{ages}} we plot the age of the protoclusters (the sink particles
that group together to form the \sfcsp) and their position from their common
centre of mass. All the points are plotted at t = 13 Myr, with the positions
being the y-direction in the simulation, since the \sfcs are more spaced out in
this direction. The ages are determined in two ways. The crosses denote the
ages determined by when the protocluster first forms. The filled hexagons are
determined by the time when the protoclusters reach a mass of 134\solmas, the
point at which a 10\solmas star can form. Note that in the previous sections we
determined when 10\solmas stars could form based on the mass contained in an
\sfc, rather than its constituent protoclusters and gas. We are forced to use
the individual protoclusters here since the \sfcs are not coherent objects
throughout the entire evolution of the simulation.

We see clearly from figure {\ref{ages}} that while a large range of ages exist
at any particular distance from the centre of mass, no trend is present in the
ages with distance. Thus our simulation predicts that the OB association would
be essentially coeval. However this is just a symptom of our idealised initial
conditions. The initial uniform density sphere, with multiple Jeans masses,
allows the entire cloud to proceed directly to star formation, via the
dissipation of kinetic energy. Since the turbulence is the same throughout the
cloud, star formation occurs simultaneously in quite separate locations. If on
the other hand our GMC needs to accumulated in a large scale shock, as
suggested by \citet{Pringleetal2001}, then there would naturally be an age
spread as the layer in which the GMC forms starts to grow. The most important
point in this picture is that the whole region would not be at the same
density, but instead would have to evolve to star forming densities as the GMC
accumulates.

\section{Conclusions}
\label{finish}

The simulation presented in this paper highlights that GMCs need not be
regarded as objects in virial equilibrium, or even bound, for them to be sites
of star formation. Globally unbound GMCs can form stellar clusters very
quickly, on roughly their crossing time. Furthermore, the unbound state of the
cloud ensures that whole region is also dispersing while it is forming stars.
They are thus naturally transient features. This evolutionary picture of a
cloud forming, producing a stellar population, and then dispersing has been
shown by Elmegreen (2000) to be apparent in a number of independent
observations.

Using some simple assumptions about the form of the star formation in the star
formation centres (\sfcsp) of our simulation, we have provided an estimate of
the star formation efficiency in the GMC. At the point one would expect the
first supernova events, we find that the star formation efficiency is about
5 - 10\%. This assumes that the \sfc environment in the simulation yields an
efficiency of \sims 50\%. Removing this assumption about the \sfcsp, and
letting the SN event be the only control over the efficiency, we find that the
cloud has a global star formation efficiency of \sims 7-8\% (for our assumed
IMF).

We argue that unbound GMCs may provide a simple mechanism for forming OB
associations, a concept that was touched upon by \citet{Ambart1955,
Ambart1958}. OB stars form at the centre of a population of \sfcsp.
These \sfcsp, which condense out of the unbound flows in the GMC are
naturally expanding away from one another, as the positive energy disperses the
cloud's gas. Not only does the mechanism explain the OB association dynamics
but it also explains the observed substructure, generally referred to as OB
subgroups. Since the OB association is just a series of independently formed
clusters, one would also expect the association to have the field star IMF.

\section*{Acknowledgments}

The computations reported here were performed using the UK Astrophysical Fluids
Facility (UKAFF). We also acknoweledge the assistance of the EC-network grant
EC-RTN1-1999-00436. The authors would like to dedicate this paper to Adriaan
Blaauw, on the occasion of his 90th birthday.

\bibliographystyle{mn2e}
\bibliography{GMC}

\end{document}